\RequirePackage{fix-cm}
\documentclass[smallextended]{svjour3}       
\smartqed  
\usepackage{graphicx}
\begin{document}

\title{A shrinkage probability hypothesis density filter for multitarget tracking
}


\author{Huisi Tong         \and
        Hao Zhang         \and
        Huadong Meng         \and
        Xiqin Wang 
}


\institute{Huisi Tong \at
              Department of Electronic Engineering,Tsinghua University \\
              Tel.: +8613466749513\\
              \email{tonghs08@mails.tsinghua.edu.cn}           
           \
}

\date{Received: date / Accepted: date}

\maketitle

\begin{abstract}
 In radar systems, tracking targets in low signal-to-noise ratio (SNR) environments is a very important task. There are some algorithms designed for multitarget tracking.   Their performances, however, are not satisfactory in low SNR environments. Track-before-detect (TBD) algorithms have been developed as a class of improved methods for tracking in low SNR environments. However, multitarget TBD is still an open issue. In this paper, multitarget TBD measurements are modeled, and a highly efficient filter in the framework of finite set statistics (FISST) is designed. Then, the probability hypothesis density (PHD) filter is applied to multitarget TBD. Indeed, to solve the problem of the target and noise not being separated correctly when the SNR is low, a shrinkage-PHD filter is derived, and the optimal parameter for shrinkage operation is obtained by certain optimization procedures. Through simulation results, it is shown that our method can track targets with high accuracy by taking advantage of shrinkage operations.
\end{abstract}

\section{Introduction}
\label{intro}

In order to extract target measurements, traditional tracking
methods apply a detection threshold at every scan. The undesirable
effect of detecting the sensor data, however, is that in restricting
the data flow, it also throws away potentially useful information.
For high signal-to-noise ratio (SNR) targets, this loss of
information is of little concern \cite{Ref1}. For low SNR targets,
this loss of information could be critical for a radar tracking
system. Therefore, some new algorithms using unthresholded data are
more advantageous than the traditional methods in tracking low SNR
targets.

The concept of simultaneous detection and tracking using
unthresholded data is known in literature as track-before-detect
(TBD) approach \cite{Ref1}. TBD algorithms could improve the
performance of a tracking system, which has been investigated for
surveillance radar \cite{Ref2}. In \cite{Ref3} and \cite{Ref4}, the
advantage of TBD methods is discussed and many TBD methods are
reviewed and compared. As a batch algorithm using the Hough
transform \cite{Ref5}, dynamic programming \cite{Ref6} or maximum
likelihood estimation \cite{Ref7}, TBD could be implemented. These
techniques operate on several data scans and in general require
large computational resources \cite{Ref1}.

As an alternative, recursive TBD method is based on a recursive
single-target Bayes filter \cite{Ref1}. An extension of the particle
filter to multitarget TBD is given in \cite{Ref8}, and an improved
approach is given in \cite{Ref9}. In this algorithm, a modeling
setup is applied to accommodate the varying number of targets. Then
a multiple model Sequential Monte Carlo based TBD approach is used
to solve the problem conditioned on the model, i.e., the number of
targets \cite{Ref10}. This approach has proven to be very efficient
in both single and multitarget case \cite{Ref3}, though it restricts
itself to the case in which the maximum possible number of targets
is limited and known.

Another extension of the single-target Bayes filter to multitarget
TBD is based on a multitarget Bayes filter. Because a single-target
Bayes filter is optimal for a single target, to solve the problems
introduced by multiple targets, the multitarget Bayes filter is
proposed in \cite{Ref11}. In a multitarget Bayes filter, multitarget
states and observations are modeled as random finite sets (RFS).
This approach is a theoretically optimal approach to multitarget
tracking in the framework of finite set statistics (FISST)
\cite{Ref12}. However, the multitarget Bayes filter has no practical
utility without an approximation strategy. To solve this problem,
the probability hypothesis density (PHD) filter \cite{Ref13}  is
proposed as a tractable and calculation-simple alternative to the
multitarget Bayes filter \cite{Ref14}.

The PHD is the first moment of the multitarget posterior
probability. Under some assumptions (e.g. Possion Assumption), the
PHD is an approximation of the multitarget posterior probability.
Therefore, the PHD filter can be an approximation of the multitarget
Bayes filter. Although a PHD algorithm for TBD is proposed in
\cite{Ref10}, this approach ignores TBD measurements should be
modeled by RFSs. In \cite{Ref15}, multitarget TBD from image
observations is formulated in a Bayesian framework by modeling the
collection of states as a Multi-Bernoulli RFS. This work use the
multi-Bernoulli update to develop a high precision multi-object
filtering algorithm for image observations, although its
adaptability of low SNR environment is needed to discussed in more
detail.

In our previous work  \cite{Ref16}, we use the RFSs to model
multitarget TBD measurements and the collection of states. In this
way, a traditional PHD filter \cite{Ref12} and \cite{Ref13} could be
applied to multitarget TBD. Even though PHD could be an
approximation of the multitarget posterior probability, the accuracy
of the algorithm is limited by some reasons, which are indicated in
this paper. Firstly, when the SNR is too low, the PHD cannot be a
sufficient approximation of the multitarget posterior probability
because the fundamental assumptions are challenged. Furthermore, for
multitarget TBD, the measurements of target and noise can hardly be
separated, while the PHD filter is heavily dependent on the
measurements of targets \cite{Ref18}. In traditional tracking
systems, to solve these problems, cardinalized PHD filter (CPHDF) is
proposed in \cite{Ref19}. However, CPHDF is inefficient in
multitarget TBD, because the computational complexity of CPHDF is
$O\left( {{m^3}} \right)$ , where  is the number of elements of the
measurement set and is quite large for the TBD problem.

In this paper, for extending a traditional PHD filter to be suitable
for multitarget TBD, viewed from a different perspective, TBD can be
regarded as a kind of classification problem of target and noise
measurements. To enhance the difference between the target and noise
measurements and to pursue better classification performance, the
measurements need to be denoised. The threshold shrinkage algorithm
\cite{Ref20} is an important method for image denoising. In general,
the key points of threshold shrinkage are the following: the method
of shrinkage (e.g., soft-threshold method) and the selection
criterion of the threshold. In this paper, a shrinkage operation is
adopted that is similar to the threshold shrinkage algorithm. The
optimal parameter for the shrinkage operation can be obtained via
certain optimization procedures.

Furthermore, in this paper, some problems of multitarget TBD in
particle use are also discussed. The assumption of known SNRs is
used frequently in traditional TBD algorithms \cite{Ref6},
\cite{Ref8} and \cite{Ref9}. Multiple targets with similar SNRs are
common assumption in the simulations of PHD algorithms using the
amplitude feature, as in \cite{Ref15} and \cite{Ref21}. However, in
practical use, multiple targets with different or unknown SNRs are
common. In this paper, for the SMC implement of the PHD filter
adopted, this problem could be solved by augmenting the SNR into
target state, varying the method of generating predicted particles
and adjusting the update operator.

In this paper, the measurement of targets is modeled by a
'nail-like' model on range-Doppler maps because of the assumptions
of the classical PHD filter in the framework of FISST. Recently, a
classical PHD filter has been modified to solve the problems of
extended and group targets in \cite{Ref22} and \cite{Ref23},
respectively. Therefore, the TBD measurement model of extended
targets in the framework of FISST will be discussed in future work.

The rest of this paper is organized as follows. In Section 2, the
multitarget TBD problem is modeled by RFSs. In Section 3, the
limitation of traditional PHD filter extension to TBD is
investigated. In Section 4, a shrinkage option for the PHD filter is
proposed with an optimal parameter. Tracking multiple targets with
different or unknown SNRs is discussed in Section 5. Simulation
results for the tracking system are presented in Section 6. Finally,
we conclude the paper in Section 7.

\section{Multitarget TBD RFS model}
\label{sec:1}
\subsection{State of the RFS model}

 The target state is ${{\bf{x}}_k} = {\left[ {{x_k},{{\dot x}_k},{y_k},{{\dot y}_k}}
\right]^T}$,where $({x_k},{y_k})$ and $({\dot x_k},{\dot y_k})$ are
the position and velocity. Since there is no ordering on the
respective collections of all target states, they can be naturally
represented as a finite set. ${X_k}$ is the multitarget state-set at
time step $k$, i.e., the set of unknown target states (which are
also of unknown number).

In a single-target scenario, the state ${{\bf{x}}_k}$ is modeled as
random vectors. Then, the multitarget state, including target
motion, birth, spawning, can be described by RFS. For target motion,
given multitarget state set ${X_{k - 1}}$, each ${{\bf{x}}_{k - 1}}
\in {X_{k - 1}}$ either survives at time step $k$ with probability
${e_{k\left| {k - 1} \right.}}({{\bf{x}}_{k - 1}})$, and its
transition probability density from ${x_{k - 1}}$ to $x_k$ is
${f_{k\left| {k - 1} \right.}}({{\bf{x}}_k}\left| {{{\bf{x}}_{k -
1}}} \right.)$ . Therefore, the target motion is modeled as the RFS
${S_{k\left| {k - 1} \right.}}({{\bf{x}}_{k - 1}})$. In the same
way, when the RFS of target birth at time $k$ is modeled by ${\Gamma
_k}$, and the RFS of targets spawning from a target with
${{\bf{x}}_{k - 1}}$ is modeled by  ${B_{k\left| {k - 1}
\right.}}({{\bf{x}}_{k - 1}})$ , the multitarget stat ${X_k}$ is
given by
\begin{equation}\label{eq-1}
{X_k} = \left[ {\bigcup\limits_{{{\bf{x}}_{k - 1}} \in {X_{k - 1}}}
{{S_{k\left| {k - 1} \right.}}({{\bf{x}}_{k - 1}})} } \right]\bigcup
{\left[ {\bigcup\limits_{{{\bf{x}}_{k - 1}} \in {X_{k - 1}}}
{{B_{k\left| {k - 1} \right.}}({{\bf{x}}_{k - 1}})} } \right]\bigcup
{{\Gamma _k}} }.
\end{equation}

\subsection{Measurement of the RFS model}
\label{sec:2} The measurements are measurements of reflected power
as in \cite{Ref8}. ${n_k}$ is white complex Gaussian noise with
variance $\sigma _0^2$. In this section, assume that the intensity
of all targets is ${I_k}$, the SNR for the targets is defined by

\begin{equation}\label{eq-2}
SNR = 10\log (\frac{{I_k^2}}{{2\sigma _0^2}}){\rm{  }}dB
\end{equation}

The method to deal with multiple targets with different or unknown
SNRs is discussed in section 5.

 Because of the assumptions made in
the modeling process and the essential difference between an RFS and
a random vector, the two conditions that should be satisfied to
model TBD measurements by a RFS are the following: there is no
target that generates more than one measurement vector, and no
measurement vector is generated by more than one target\cite{Ref22}.
In summary, the measurement of targets should be modeled by a
'nail-like' model on the range-Doppler-Bearing maps (the $ijk$ cell
is defined by coordinat ($r_i,d_j,b_l$)) as follow:

R, D and B are the size of a range, the Doppler and the bearing
cell. $r_k = \sqrt {{{(x_k)}^2} + {{(y_k)}^2}}$, $d_k =
\frac{{x_k\dot x_k+ y_k\dot y_k}}{{\sqrt {{{(x_k)}^2} + {{(y_k)}^2}}
}}$ and $b_k = \arctan \left( {\frac{{y_k}}{{x_k}}} \right)$. $R_k^i
= \left[ {{r_i} - \frac{R}{2},{r_i} + \frac{R}{2}} \right)$, $D_k^j
= \left[ {{d_j} - \frac{D}{2},{d_j} + \frac{D}{2}} \right)$ and
$B_k^l = \left[ {{b_l} - \frac{B}{2},{b_l} + \frac{B}{2}} \right)$.

Then the measurement model is
\begin{equation}\label{eq-3}
z_k^{ijl} = {\left| {{h^{ijl}}({{\bf{x}}_k}) + {n_k}} \right|^2}
\end{equation}

\begin{equation}\label{eq-4}
h^{ijl} = \left\{ \begin{array}{ll}{I_k} & \textrm{if ${{r_k} \in
R_k^i,{d_k} \in D_k^jand{b_k} \in B_k^l}$}
\\0 &\textrm{else}\end{array} \right.
\end{equation}

This means the measurement of a target is like a nail on the
range-Doppler-bearing maps. This 'nail-like' model is similar to the
point target model in \cite{Ref5} and \cite{Ref6}. At step k, the
measurement provided by the sensor consists of
 $N = {N_r} \times {N_d} \times {N_b}$ measurements $z_k^{ijl}$, where  ${N_r}$, ${N_d}$  and ${N_b}$ are the number of range,
Doppler and bearing cells.

The number of $z_k^{ijl}$ is a constant $N$. However, the number of
element of an RFS should be random and Poisson distributed for a PHD
filter. Because weak signal information should be preserved by TBD
algorithms, we should make sure all measurements produced by targets
are included in the RFS. Therefore, a threshold could be chosen as
follows:
\begin{equation}\label{eq-5}
\begin{array}{l}
 {\theta _k} = \mathop {\arg \min }\limits_{{\theta _k}} \int_{{\theta _k}}^{ + \infty } {{g_0}(z_k^{ijl})} dz_k^{ijl} \\
 s.t.\begin{array}{*{20}{c}}
   {}  \\
\end{array}{p_D}({{\bf{x}}_k}) ={P(z_k^{ijl} \ge {\theta _k})} =\int_{{\theta _k}}^{ + \infty } {{g_k}(z_k^{ijl}\left| {{{\bf{x}}_k}} \right.)} dz_k^{ijl} \ge 0.99 \\
 \end{array}
 \end{equation}
 where the likelihood function for the target is
\begin{equation}\label{eq-6}
 {g_k}(z_k^{ijl}\left| {{{\bf{x}}_k}} \right.) = \frac{1}{{2\sigma _0^2}}\exp \left\{ { - \frac{{z_k^{ijl} + I_k^2}}{{2\sigma _0^2}}} \right\}{{\rm I}_0}\left( {\frac{{{I_k}\sqrt {z_k^{ijl}} }}{{\sigma _0^2}}} \right)
 \end{equation}
 and that for noise is
\begin{equation}\label{eq-7}
{p_0}(z_k^{ijl}) = \frac{1}{{2\sigma _0^2{\rm{ }}}}\exp \left\{ { -
\frac{1}{{2\sigma _0^2{\rm{ }}}}z_k^{ijl}}
\right\}.\begin{array}{*{20}{c}}
   {}  \\
\end{array}
 \end{equation}

where ${{\rm I}_0}\left( {} \right)$ is the zero order Bessel
function and $z_k^{ijl}$ is assumed positive.The
${p_D}({{\bf{x}}_k})$ is the probability detection. Because this
algorithm is for TBD, it should be ensured that
${p_D}({{\bf{x}}_k})\approx 1$.

$Z_k$ is the observation-set consisting of all measurements
collected by all sensors at time-step $k$, no matter the measurement
from the targets or from the noise. If $z_k^{ijl} \ge {\theta _k}$,
let ${z_k} = z_k^{ijl}$. Then, the measurement RFS $Z_k$ is
constructed by a subset of the $z_k^{ijl}$
 using a thresholding mechanism. The threshold can insure that ${p_D}({{\bf{x}}_k})\approx 1$ by (\ref{eq-5}). The threshold is a function of the SNR of the targets. The element in the RFS $Z_k$ is as follows:
\begin{equation}\label{eq-8}
{\bf{z}}_k^*{\rm{ = }}{\left[ {{r_i},{d_j},{b_l},{z_k}} \right]^T}
\end{equation}

and the likelihood function according
\begin{equation}\label{eq-d1}
 g_k^*({\bf{z}}_k^*\left| {{{\bf{x}}_k}} \right.)\approx {g_k}(z_k\left| {{{\bf{x}}_k}} \right.)
\end{equation}

\begin{equation}\label{eq-d2}
p_0^*({\bf{z}}_k^*) = \frac{1}{{2{\sigma_0 ^2}{\rm{ }}}}\exp \left\{
{ - \frac{{{z_k} - {\theta _k}}}{{2{\sigma_0 ^2}{\rm{ }}}}} \right\}
\end{equation}
At the same time, the sensor could be modeled by
\begin{equation}\label{eq-9}
{Z_k} = {K_k}\bigcup {\left[ {\bigcup\limits_{{{\bf{x}}_k} \in
{X_k}} {{\Theta _k}({{\bf{x}}_k})} } \right]}, {\bf{z}}_k^* \in
{Z_k}
\end{equation}
where the measurements of targets are modeled by RFS ${\Theta
_k}({x_k})$ and the noise is modeled as RFS ${K_k}$. The elements in
${\Theta _k}({x_k})$ are the measurements produced by the targets.
The elements in ${K_k}$ are the measurements which are produced by
the noise and bigger than ${\theta_k}$ as well. It is shown that
when the measurements of TBD are modeled by RFSs, multitarget TBD
can be regarded as a kind of classification problem of target and
noise measurements for one scan. This classification problem will be
further analyzed in Section 4.

\section{The traditional PHD filter extension to TBD}
\label{sec:3}

After multitarget TBD measurements and the collection of states are
modeled by RFS, a traditional PHD filter is applied to multitarget
TBD. This algorithm is reviewed in 3.1.  However, the accuracy of
the algorithm is limited by some reasons when the SNR is low, which
is discussed in Section 3.2.

\subsection{The algorithm}
\label{sec:4} As proposed in \cite{Ref13}, the PHD prediction and
update equations are presented as follows:

\begin{equation}\label{eq-tex}
\begin{array}{*{20}{c}}
{{D_{k\left| {k{\rm{ - 1}}} \right.}}({\bf{x}}\left| {{Z_{1:{k-1}}}}
\right.)  = \int {{e_{k\left| {k - 1} \right.}}(\zeta ){f_k}({\bf{x}}\left| \zeta  \right.){D_{k - 1\left| {k{\rm{ - 1}}} \right.}}(\zeta \left| {{Z_{1:k - 1}}} \right.)d\zeta  + } }\\
{\int {{b_{k\left| {k - 1} \right.}}({\bf{x}}\left| \zeta
\right.){D_{k - 1\left| {k{\rm{ - 1}}} \right.}}(\zeta \left|
{{Z_{1:k - 1}}} \right.)d\zeta } {\rm{ + }}{\gamma
_k}({\bf{x}}){\rm{  }}}
\end{array}
\end{equation}

\begin{equation}\label{eq-b2}
{D_{k\left| k \right.}}({\bf{x}}\left| {{Z_{1:k}}} \right.) = \left[
{{\rm{1 - }}{p_D}({\bf{x}})} \right]{D_{k\left| {k{\rm{ - 1}}}
\right.}}({\bf{x}}) + \sum\limits_{{z_k} \in {Z_k}}^{}
{\frac{{{p_D}({\bf{x}}){g_k}({{\bf{z}}_k}\left| {\bf{x}}
\right.){D_{k\left| {k{\rm{ - 1}}} \right.}}({\bf{x}})}}{{{\kappa
_k}({{\bf{z}}_k}) + \int {{p_D}(\zeta ){g_k}({{\bf{z}}_k}\left|
\zeta  \right.){D_{k\left| {k{\rm{ - 1}}} \right.}}(\zeta )d\zeta }
}}}
\end{equation}
where ${D_{k\left| k \right.}}({\bf{x}}\left| {{Z_{1:k}}} \right.)$
the PHD is the density whose integral $\int_S {{D_{k\left| k
\right.}}({\bf{x}}\left| {{Z_{1:k}}} \right.)d{\bf{x}}}$
 on any region S of state space is $\hat n_k^{}(S) = \int {\left| {X \cap S} \right|} {p_k}({X_k}\left| {{Z_{1:k}}} \right.)\delta
 X$. ${b_{k\left| {k - 1} \right.}}({{\bf{x}}_k}\left| {{{\bf{x}}_{k -
1}}} \right.)$ and ${\gamma _k}({{\bf{x}}_k}){\rm{ }}$ denotes the
intensity of ${B_{k\left| {k - 1} \right.}}({{\bf{x}}_{k - 1}})$ and
${\Gamma _k}$ at time $k$, and ${\kappa _k}({{\bf{z}}_k})$ is the
intensity of ${K_k}$. $Z_{1:k}$ is the time-sequence of
observation-sets.

Because the PHD propagation equations involve multiple integrals
that have no computationally tractable closed form expressions,
Sequential Monte Carlo (SMC) methods are used to approximate the PHD
in \cite{Ref12}. Let the update PHD at $k-1$ step ${D_{k - 1\left|
{k{\rm{ - 1}}} \right.}}({\bf{x}}\left| {{Z_{1:k - 1}}} \right.)$ be
represented by a set of particles $\left\{ {\omega _{k -
1}^{(p)},{\bf{x}}_{k - 1}^{(p)}} \right\}_{p = 1}^{{L_{k - 1}}}$ ,
as

\begin{equation}\label{eq-a1}
{D_{{k-1}\left| {k-1} \right.}}({\bf{x}}\left| {{Z_{1:k}}} \right.)
= \sum\limits_{p = 1}^{{L_{k - 1}}} {\omega _{k-1}^{*(p)}\delta
({\bf{x}} - {\bf{x}}_{k - 1}^{(p)})}
\end{equation}

The predicted particles are generated by

\begin{equation}\label{eq-b2}
{\bf{x}}_{\left. k \right|k - 1}^{(p)} \sim \left\{
\begin{array}{l}
 {q_k}( \bullet \left| {{\bf{x}}_{k - 1}^{(p)}} \right.)\begin{array}{*{20}{c}}
   {} & {} & {}  \\
\end{array}p = 1, \ldots ,{L_{k - 1}} \\
 {v_k}( \bullet ){\rm{  }}\begin{array}{*{20}{c}}
   {} & {}  \\
\end{array}{\rm{ }}p = {L_{k - 1}} + 1, \ldots ,{L_{k - 1}} + {J_k} \\
 \end{array} \right.{\rm{ }}
\end{equation}

where ${q_k}( \bullet \left| {{\bf{x}}_{k - 1}^{(p)}} \right.)$ and
${v_k}( \bullet )$ are proposal density. The predicted density is

\begin{equation}\label{eq-b3}
{D_{k\left| {k - 1} \right.}}({\bf{x}}\left| {{Z_{1:k - 1}}}
\right.) = \sum\limits_{p = 1}^{{L_{k - 1}}{\rm{ + }}{J_k}} {\omega
_{\left. k \right|k - 1}^{(p)}\delta ({\bf{x}} - {\bf{x}}_{\left. k
\right|k - 1}^{(p)})}
\end{equation}

where

\begin{equation}\label{eq-b4}
\omega _{\left. k \right|k - 1}^{(p)} = \left\{ \begin{array}{r}
 \frac{{{e_{k\left| {k - 1} \right.}}({\bf{x}}_{k - 1}^{(p)}){f_k}({\bf{x}}_{\left. k \right|k - 1}^{(p)}\left| {{\bf{x}}_{k - 1}^{(p)}} \right.) + {b_{k\left| {k - 1} \right.}}({\bf{x}}_{\left. k \right|k - 1}^{(p)}\left| {{\bf{x}}_{k - 1}^{(p)}} \right.)}}{{{q_k}({\bf{x}}_{\left. k \right|k - 1}^{(p)}\left| {{\bf{x}}_{k - 1}^{(p)}} \right.)}}{\rm{   }} \\
 p = 1, \ldots ,{L_{k - 1}} \\
 \frac{{{\gamma _k}({\bf{x}}_{\left. k \right|k - 1}^{(p)})}}{{{v_k}({\bf{x}}_{\left. k \right|k - 1}^{(p)})}}{\rm{        }}p = {L_{k - 1}} + 1, \ldots ,{L_{k - 1}} + {J_k} \\
 \end{array} \right.
\end{equation}

Set ${p_D}({\bf{x}}) \equiv 1$, the update density will be

\begin{equation}\label{eq-10}
{D_{k\left| k \right.}}({\bf{x}}\left| {{Z_{1:k}}} \right.) =
\sum\limits_{p = 1}^{{L_{k - 1}} + {J_k}} {\omega _k^{*(p)}\delta
({\bf{x}} - {\bf{x}}_{\left. k \right|k - 1}^{(p)})}
\end{equation}

where

\begin{equation}\label{eq-b5}
\omega _k^{*(p)} = \sum\limits_{{\bf{z}}_k^{} \in {Z_k}}^{}
{\frac{{{g_k^*}({\bf{z}}_k^*{}\left| {{\bf{x}}_{\left. k \right|k -
1}^{(p)}} \right.)\omega _{k\left| {k - 1} \right.}^{(p)}}}{{{\kappa
_k}({\bf{z}}_k^{}) + \sum\limits_{p = 1}^{{L_{k - 1}} + {J_k}}
{{g_k^*}({\bf{z}}_k^*{}\left| {{\bf{x}}_{\left. k \right|k -
1}^{(p)}} \right.)\omega _{k\left| {k - 1} \right.}^{(p)}} }}}
\end{equation}

According to the standard treatment of particle filter, resample
$\left\{ {\omega _k^{*(p)}/\hat n_k^{},{\bf{x}}_{\left. k \right|k -
1}^{(p)}} \right\}_{p = 1}^{{L_{k - 1}} + {J_k}}$ to get $\left\{
{\omega _k^{(p)}/\hat n_k^{},{\bf{x}}_k^{(p)}} \right\}_{p =
1}^{{L_k}}$

\subsection{The limitation}
\label{sec:5}

Because ${\kappa _k}({\bf{z}}_k^{})$ can be presented by the number
of noise samples ${\lambda _k}$ multiplying the clutter probability
density, the update operator (\ref{eq-b5}) turns into

\begin{equation}\label{eq-11}
\omega _k^{*(p)} = \sum\limits_{{\bf{z}}_k^* \in {Z_k}}^{}
{\frac{{{g_k^*}\left( {{\bf{z}}_k^*\left| {{\bf{x}}_{\left. k
\right|k - 1}^{(p)}} \right.} \right)}}{{\lambda_k p_0^*\left(
{{z_k}:{\sigma _0}} \right) + \sum\limits_{p = 1}^{{L_{k - 1}} +
{J_k}} {{g_k}^*\left( {{\bf{z}}_k^*\left| {{\bf{x}}_{\left. k
\right|k - 1}^{(p)}} \right.} \right)\omega _{k\left| {k - 1}
\right.}^{(p)}} }}} \omega _{k\left| {k - 1} \right.}^{(p)}.
\end{equation}
\begin{equation}\label{eq-12}
p_0^*(z_k:\sigma) = \frac{1}{{2{\sigma ^2}{\rm{ }}}}\exp \left\{ { -
\frac{{{z_k} - {\theta _k}}}{{2{\sigma ^2}{\rm{ }}}}} \right\}
\end{equation}
Equation (\ref{eq-12}) is the likelihood function for the
measurement of noise in the measurements set, which denotes the
clutter probability density in theoretical filtering (\cite{Ref12}
and \cite{Ref13}), and the optimal choice of  $\sigma$ should be the
variance of noise , as shown in (\ref{eq-11}). However, for TBD
applications, the SNR is extremely low. Hence, this setting of leads
to the situation in which measurements generated by the target and
noise are not separated correctly and heavy degradation of tracking
performance is seen. In fact, the derivation of (\ref{eq-11})
depends on the hypothesis that the number of noise sample is Poisson
distributed. This means the clutter follows a Poisson distribution
and is independent of target-originated measurements. In this way,
the PHD can be the "best fit" approximations of the multitarget
posterior probability \cite{Ref13}. The average number of noise
samples included in RFS is

\begin{equation}\label{eq-13}
\lambda_k  = N\exp \left\{ { - \frac{{{\theta _k}}}{{2\sigma _0^2}}}
\right\}.
\end{equation}

When $N$ is large and $\lambda_k$ is relatively small, the Poisson
distribution could be approximated by the distribution of a number
of noise samples. However, this approximation will be destroyed when
become sufficiently large, which is the case in low SNR scenarios,
as in Table 1.

\begin{table}[!h]
\caption{$\lambda $versus SNR (N = 2000)}
\label{tab:1}       
\begin{tabular}{llllll}
\hline\noalign{\smallskip}
SNR(dB) & 6 & 7 & 8 & 9 & 10 \\
\noalign{\smallskip}\hline\noalign{\smallskip}
$\lambda_k $ & 1340 & 1098 & 707 & 344 & 143 \\
\noalign{\smallskip}\hline
\end{tabular}
\end{table}

When  ${\sigma _0}$ in  (\ref{eq-11}) cannot make PHD a sufficient
approximation of the multitarget posterior probability, choosing the
"optimal" becomes the key problem for extending the PHD filter to be
applied in TBD. As discussed in the following section, this
selection can be considered as a shrinkage operation, and the
optimal parameter for the shrinkage operation can be obtained via
certain optimization procedures.

\section{Shrinkage operation for TBD}
\label{sec:4} To determine the "optimal"   and make the noise and
targets distinguishable, a different perspective should be taken. As
referred to in Section 2.2, TBD can be regarded as a kind of
classification problem for target and noise measurements. In other
words, when a measurement is obtained, should it be classified into
a target class or a noise class? To solve this problem, the Fisher
class separability criterion, as a supplement of to the traditional
Bayesian framework, was adopted in our design for the extension of
the PHD filter. Furthermore, to enhance the difference between the
target and noise measurements and pursue the better performance of
classification, some attempts to reduce the noise are needed. The
threshold shrinkage algorithm \cite{Ref20} is an important method
for the image denoising. In general, the cores of threshold
shrinkage are as follows: the method of shrinkage and the selection
criterion of the threshold. A shrinkage operation is adopted in this
section that is similar to the threshold shrinkage algorithm to some
extent. The optimal parameter for the shrinkage operation, which
acts as the threshold, can be obtained via certain optimization
procedures. The PHD filter with the shrinkage operation is called
the Shrinkage-PHD filter.

\subsection{Fisher class separability criterion}
It is known that the Bayesian classifier is the optimal classifier
when the posterior probability can be calculated. However, as
indicated in Section 3.2, the PHD cannot be a sufficient
approximation of multitarget posterior probability when the SNR of
the targets is low. Therefore, another classification criterion, the
Fisher class separability criterion, is introduced into our methods.

The separation between the two classes of objects defined by Fisher
is the ratio of the distance between the centers of classes to the
scattering of the classes \cite{Ref24}:

\begin{equation}\label{eq-14}
S = \frac{{d_{between}^2}}{{d_{within}^2}} = \frac{{{{({\mu _1} -
\mu _0^*)}^2}}}{{d_1^2 + d{{_0^*}^2}}}.
\end{equation}

where ${\mu _i}$  represents the centers of classes and $d_i^2$
denotes the scattering of the classes. In our analysis, they are set
to the mean and variance of the likelihood functions of the target
and the noise, respectively.

For targets, the mean and variance are defined as follows:

\begin{equation}\label{eq-15}
{\mu _1} = \int_\theta ^\infty  {{z_k}{g_k}({z_k}\left|
{{{\bf{x}}_k}} \right.)} d{z_k} \approx \int_0^\infty
{{z_k}{g_k}({z_k}\left| {{{\bf{x}}_k}} \right.)} d{z_k} = 2\sigma
_0^2 + I_k^2
\end{equation}

\begin{equation}\label{eq-16}
d_1^2 = \int_\theta ^\infty  {z_k^2{g_k}({z_k}\left| {{{\bf{x}}_k}}
\right.)d{z_k}}  \approx \int_0^\infty  {z_k^2{g_k}({z_k}\left|
{{{\bf{x}}_k}} \right.)} d{z_k} = 4\sigma _0^2(\sigma _0^2 + I_k^2)
\end{equation}

For the noise, the mean and variance are given by

\begin{equation}\label{eq-17}
\mu _0^* = {\mu _0} + {\theta _k} = 2\sigma _0^2 + {\theta _k}
\end{equation}

\begin{equation}\label{eq-18}
d{{_0^*}^2} = d_0^2 = 4\sigma _0^4
\end{equation}

Therefore, the Fisher class separability is

\begin{equation}\label{eq-19}
S = \frac{{{{(2\sigma _0^2 + I_k^2 - \theta  - 2\sigma
_0^2)}^2}}}{{4\sigma _0^2(\sigma _0^2 + I_k^2) + 4\sigma _0^4}} =
\frac{{{{\left( {\frac{{I_k^2}}{{2\sigma _0^2}} - \frac{{{\theta
_k}}}{{2\sigma _0^2}}} \right)}^2}}}{{2\left( {1 +
\frac{{I_k^2}}{{2\sigma _0^2}}} \right)}}.
\end{equation}

\begin{figure}[!h]
  \includegraphics[width=0.75\textwidth]{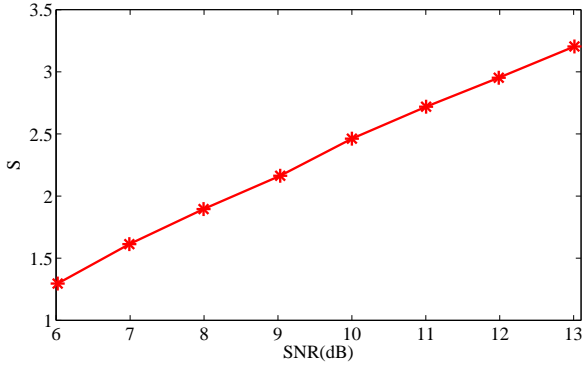}
\caption{Fig.1 The Fisher class separability versus the SNR}
\label{fig:1}       
\end{figure}

Simulations indicated that the correlation between $S$ and the SNR
of the targets is a positive, approximately linear relationship (as
shown in Fig.\ref{fig:1}). Therefore, when the SNR is low, the
Fisher class separability is so small that the targets and false
alarms could not be distinguished. In other words, we should enlarge
the Fisher class separability of the target class and the noise
class when the SNR of the targets is low.

\subsection{Shrinkage operation}

To enlarge the Fisher class separability between targets and noise,
some kind of "shrinkage" operation is applied to the likelihood
function of the noise. That is, the parameter $\sigma _0^2$
  in (\ref{eq-17}) and
(\ref{eq-18}) is adjusted. If ${\sigma _s} \le {\sigma _0}$
 is chosen instead of $\sigma _0^2$  itself, then
Fisher class separability becomes

\begin{equation}\label{eq-20}
{S_s} = \frac{{{{(2\sigma _0^2 + I_k^2 - \theta  - 2\sigma
_s^2)}^2}}}{{4\sigma _0^2(\sigma _0^2 + I_k^2) + 4\sigma _s^4}}
\end{equation}

Obviously, ${S_s}$
 is larger than $S$  whenever ${\sigma _s} \le {\sigma _0}$
. The following question naturally arises: how does one choose the
"optimal" ${\sigma _s}$ , which is the key parameter for the
shrinkage process?

\subsection{The optimal parameter for the shrinkage operation}

To demonstrate how  influences the classification problem of target
and noise measurements, we can define the Mahalanobis distance
\cite{Ref24} of measurement $z_k$ using the center of the noise
class and the target class, respectively:

\begin{equation}\label{eq-21}
{S_{s0}}({z_k},{\sigma _s}) = \frac{{{{\left( {{z_k} - {\theta _k} -
2\sigma _s^2} \right)}^2}}}{{4\sigma _s^4}}
\end{equation}

and

\begin{equation}\label{eq-22}
{S_{s1}}({z_k}) = \frac{{{{\left( {{z_k} - 2\sigma _0^2 - I_k^2}
\right)}^2}}}{{4\sigma _0^2(\sigma _0^2 + I_k^2)}}
\end{equation}

Because the selection of the threshold is a difficult task for the
threshold shrinkage algorithm \cite{Ref20} in the field of image
denoising, the selection of the optimal parameter for the
shrinkage-PHD filter involves similar problems. For the threshold
shrinkage algorithm, an oversized threshold may result in the loss
of signal. In contrast, too much noise may remain if an undersized
threshold is used. In the same way, for the shrinkage-PHD filter,
preserving the weak target information and reducing the noise by
tracking are the major challenges to be addressed for TBD problems.
Therefore, when considering the two sides, the selection criterion
should be designed as follows:

\begin{equation}\label{eq-23}
\begin{array}{l}
 \sigma _s^M = \max \begin{array}{*{20}{c}}
   {}  \\
\end{array}{\sigma _s} \\
 \begin{array}{*{20}{c}}
   {} & {}  \\
\end{array}s.t.{\rm{    }}\left\{ {\begin{array}{*{20}{c}}
   {{\sigma _s} \le {\sigma _0}}  \\
   {\int_0^{{z_s}} {{g_k}(z_k\left| {{{\bf{x}}_k}} \right.)} d{z_k} = \beta }  \\
   {{S_{s0}}({z_s},{\sigma _s}) > {S_{s1}}({z_s})}  \\
\end{array}} \right.{\rm{  }} \\
 \end{array}
\end{equation}

The first constraint represents the "shrinkage". To preserve the
weak target information and classify as many low SNR targets to the
target class as possible, the Mahalanobis distance of measurement
with the center of noise class should be enhanced; hence, the
variance of the noise class should be reduced.

The second constraint refers to the fixed probability of target
loss. The reason for this term is that for TBD problems, false
alarms can be reduced by integration over time. Meanwhile, if target
loss occurred, the integration of the information of targets is
broken off. Hence, the cost of target loss is much larger than the
cost of false alarms, which is the main difference between TBD and
common radar detection. Moreover, for the PHD filter, a missed
detection can result in loss of the track \cite{Ref18}. Therefore,
target loss should be fixed as $\beta$. When the $\beta$ is set, a
critical value $z_s$ is determined.

\begin{figure}
\begin{tabular}{c}
\includegraphics[width=0.75\textwidth]{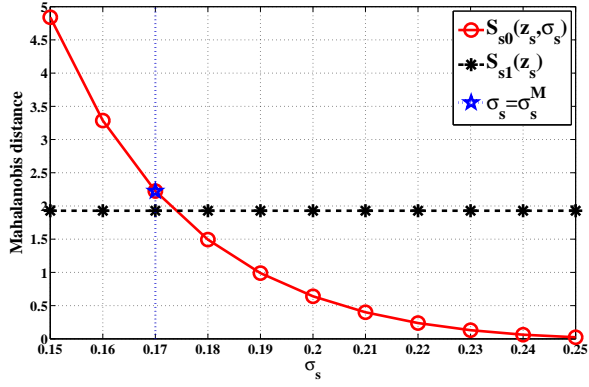}\\
(a)SNR=8dB\\
\\
\includegraphics[width=0.75\textwidth]{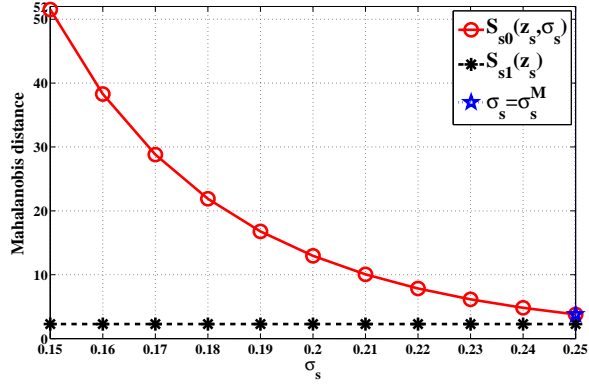}\\
(b)SNR=13dB\\
\end{tabular}
\caption{The Mahalanobis distance versus $\sigma _s$}
\label{fig:2}
\end{figure}

The third constraint means that the measurements  $z_k$ determined
by the second constraint should be classified as targets. When the
SNR is low, ${S_{s0}}({z_s},{\sigma _s})$ always intersects
${S_{s1}}({z_s})$
 , so the optimal solution $\sigma _s^M
$ to (\ref{eq-23}) is deduced, as indicated in Fig.\ref{fig:2}(a).
With the increasing SNR, ${S_{s0}}({z_s},{\sigma _s})$ becomes much
larger than ${S_{s1}}({z_s})$ and thus we choose $\sigma _s^M =
{\sigma _0}$, as shown in Fig.\ref{fig:2}(b).

Note that although the shrinkage operation could highlight the
target measurement and reduce the possibility of losing the target
during tracking, the noise might be incorrectly increased
unavoidably. To minimize the influence of shrinkage on the
classification of the noise measurement, the maximum $\sigma _s$ ,
which is consistent with the above constraints, is obtained. The
optimal solutions to (\ref{eq-23}) under different SNRs are listed
in Table 2:

\begin{table}[!h]
\caption{$\sigma _s^M$ ($\beta  \le 0.05$, ${\sigma _0} = 0.25$)}
\label{tab:2}
\begin{tabular}{lllll}
\hline\noalign{\smallskip}
SNR(dB) & 6 & 7 & 8 & 9 \\
\noalign{\smallskip}\hline\noalign{\smallskip}
$\sigma _s^M$ &$0.48{\sigma _0}$& $0.6{\sigma _0}$ & $0.68{\sigma _0}$ & $0.72{\sigma _0}$ \\
\hline\noalign{\smallskip}
SNR(dB) & 10 & 11 & 12 & 13 \\
\noalign{\smallskip}\hline\noalign{\smallskip}
$\sigma _s^M$ &$0.76{\sigma _0}$& $0.88{\sigma _0}$ & $0.96{\sigma _0}$ & ${\sigma _0}$ \\
\noalign{\smallskip}\hline
\end{tabular}
\end{table}

The update operator of the shrinkage-PHD filter can be summarized as
follows:

\begin{equation}\label{eq-c1}
\omega _{k,s}^{*(p)} = \sum\limits_{{\bf{z}}_k^*{} \in {Z_k}}^{}
{\frac{{{g_k^*}({\bf{z}}_k^*{}\left| {{\bf{x}}_{\left. k \right|k -
1}^{(p)}} \right.)\omega _{k\left| {k - 1} \right.}^{(p)}}}{{\lambda
p_0^*({z_k}:\sigma _s^M) + \sum\limits_{p = 1}^{{L_{k - 1}} + {J_k}}
{{g_k^*}({\bf{z}}_k^*{}\left| {{\bf{x}}_{\left. k \right|k -
1}^{(p)}} \right.)\omega _{k\left| {k - 1} \right.}^{(p)}} }}}
\end{equation}

where $\sigma _s^M$ should be set according to Table 2.

\section{Multitarget TBD in practical use}
This section is to solve of the problem of tracking multiple targets
with different or unknown SNRs.

As proposed in Section 4, the optimal parameter for the shrinkage
operation is closely related to the SNRs of the target, and the SNRs
of the target are assumed to be known and similar. However, in
practical use, multiple targets with different or unknown SNRs are
common. The PHD is the first moment of the multitarget posterior
probability, but not the posterior probability of a certain target.
Therefore, the ${I_k}$ in (\ref{eq-6}) and the $\sigma_s^M$ in Table
\ref{tab:2} are difficult to determine. Because the SMC implement of
the PHD filter is adopted, this problem could be solved by adjusting
the method of generating predicted particles and the update
operator.

To describe how to deal with the different SNRs of multiple targets,
we augment ${I_k}$  into ${{\bf{x}}_k}$ :
\begin{equation}\label{eq-24}
{{\bf{x}}_k} = {\left[ {{{{\bf{\tilde x}}}_k},{I_k}} \right]^T}
\end{equation}
where ${{\bf{\tilde x}}_k} = {\left[ {{x_k},{{\dot
x}_k},{y_k},{{\dot y}_k}} \right]}$. The particle of the target
state is ${\bf{x}}_k^{(p)} = {\left[ {{\bf{\tilde
x}}_k^{(p)},I_k^{(p)}} \right]^T}$. The ${\bf{\tilde x}}_{\left. k
\right|k - 1}^{(p)}$ is generated as (\ref{eq-b2})

\begin{equation}\label{eq-25}
{\bf{\tilde x}}_{\left. k \right|k - 1}^{(p)} \sim \left\{
\begin{array}{l}
 {q_k}( \bullet \left| {{\bf{\tilde x}}_{k - 1}^{(p)}} \right.)\begin{array}{*{20}{c}}
   {} & {} & {}  \\
\end{array}p = 1, \ldots ,{L_{k - 1}} \\
 {v_k}( \bullet )\begin{array}{*{20}{c}}
   {} & {}  \\
\end{array}p = {L_{k - 1}} + 1, \ldots ,{L_{k - 1}} + {J_k} \\
 \end{array} \right.{\rm{ }}
\end{equation}

The $I_k^{(p)}$ is generated by the assumption of a uniform
distribution in the range of ${{\bf{\tilde I}}_k}$, where the SNRs
of all of the targets are $\left[ {SN{R_{1,}}SN{R_{2,}} \cdots
,SN{R_m}} \right]$, and corresponding to ${{\bf{\tilde I}}_k} =
\left[ {{I_{(k,1)}}_,{I_{(k,2)}}, \cdots ,{I_{(k,m)}}} \right]$ by:

\begin{equation}\label{eq-res1}
SNR_m = 10\log (\frac{{I_{(k,m)}^2}}{{2\sigma _0^2}}){\rm{ }}dB
\end{equation}

In the update operator, because $\sigma _s^M$ is a function of the
SNR, it becomes $\sigma _s^M(I_k^{(p)})$ in (\ref{eq-c1}).

Note that the ${\theta _k}$ in (\ref{eq-5}) should also be adjusted.
Because weak signal information should be preserved by TBD
algorithms, the ${I_k}$ in (\ref{eq-6}) is the minimum of
${{\bf{\tilde I}}_k}$.

For the targets with unknown SNRs, it is assumed that the range of
the SNRs of the targets is $[SN{R_l},SN{R_h}]$, corresponding to
$[{I_{(k,l)}},{I_{(k,h)}}]$. The $I_k^{(p)}$ could be normalized for
the region $[{I_{(k,l)}},{I_{(k,h)}}]$. The update operator and
threshold are similar to those of multiple targets with different
SNRs.

\section{Simulation}
\subsection{Multiple targets miss distance}

The optimal sub-pattern assignment (OSPA) distance \cite{Ref15}
between the estimated and true multitarget state is adopted here to
estimate error.

The OSPA distance is defined as follows \cite{Ref15}. Let
${d^{(c)}}({\bf{q}},{\bf{y}}): = \min (c,\left\| {{\bf{q}} -
{\bf{y}}} \right\|)$ for ${\bf{q}},{\bf{y}} \in {\Re ^d}$, and ${\Pi
_p}$  denote the set of permutations on $\left\{ {1,2, \ldots ,p}
\right\}$ for any positive integer $p$. $c$ is the cut-off value.
Then, for $Q = \left\{ {{{\bf{q}}_1}, \cdots ,{{\bf{q}}_m}}
\right\}$ and $Y = \left\{ {{{\bf{y}}_1}, \cdots ,{{\bf{y}}_n}}
\right\}$,

\begin{equation}\label{eq-26}
{\bar d^{(c)}}\left( {Q,Y} \right): = \frac{1}{n}\left( {\mathop
{\min }\limits_{\pi  \in {\Pi _n}} \sum\limits_{i = 1}^m
{{d^{(c)}}\left( {{\bf{q}_i},{\bf{y}_{\pi (i)}}} \right) + c *
\left( {n - m} \right)} } \right)
\end{equation}
if $m \le n$; and${\bar d^{(c)}}\left( {Q,Y} \right): = {\bar
d^{(c)}}\left( {Y,Q} \right)$ if $m > n$; and ${\bar d^{(c)}}\left(
{Q,Y} \right) = {\bar d^{(c)}}\left( {Y,Q} \right) = 0$ if $m = n =
0$.

For example, when $m=1$ and $n=2$, $Q_1=\left\{q_1\right\}$ and
$Y_1=\left\{y_1,y_2\right\}$, the ${\bar d^{(c)}}\left( {Q_1,Y_1}
\right)=0.5*(min(min(|q_1-y_1|,c),min(|q_1-y_2|,c))+c)$.

\subsection{Multiple targets with the same SNR }

Consider range cells in the interval $[80000,90000]$  with meters as
the unit and Doppler cells in the interval $[-400,-150]$ with
meters/second as the unit. ${N_r} = 200$, ${N_d} = 10$ and ${N_b} =
1$. The size of a range and Doppler cell is as follows: R=50 m and
D=25 m/s. ${L_k} \equiv 2000$ and ${J_k} \equiv 800$ . The time
interval T is equal to 1 s.

At time step 1, the first target appears with the initial target
state ${\left[ {\begin{array}{*{20}{c}}
   {89000} & { - 200} & 0 & 0 \\
\end{array}} \right]^T}$. At step 10, a secondary target spawns from the first target,
moving at -300 m/s in the x-direction. The standard deviation of the
noise is 0.25.

\begin{figure}[!h]
  \includegraphics[width=0.75\textwidth]{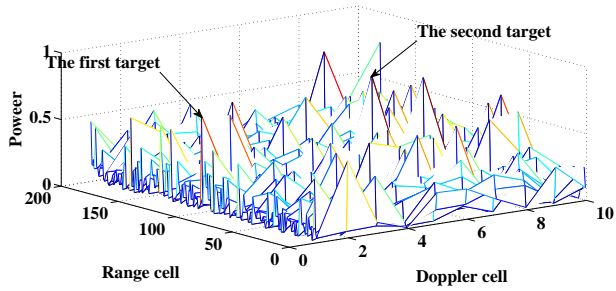}
\caption{Power measurements as a function of range and Doppler cells
at time step 20.}
\label{fig:3}       
\end{figure}

\begin{figure}[!h]
  \includegraphics[width=0.75\textwidth]{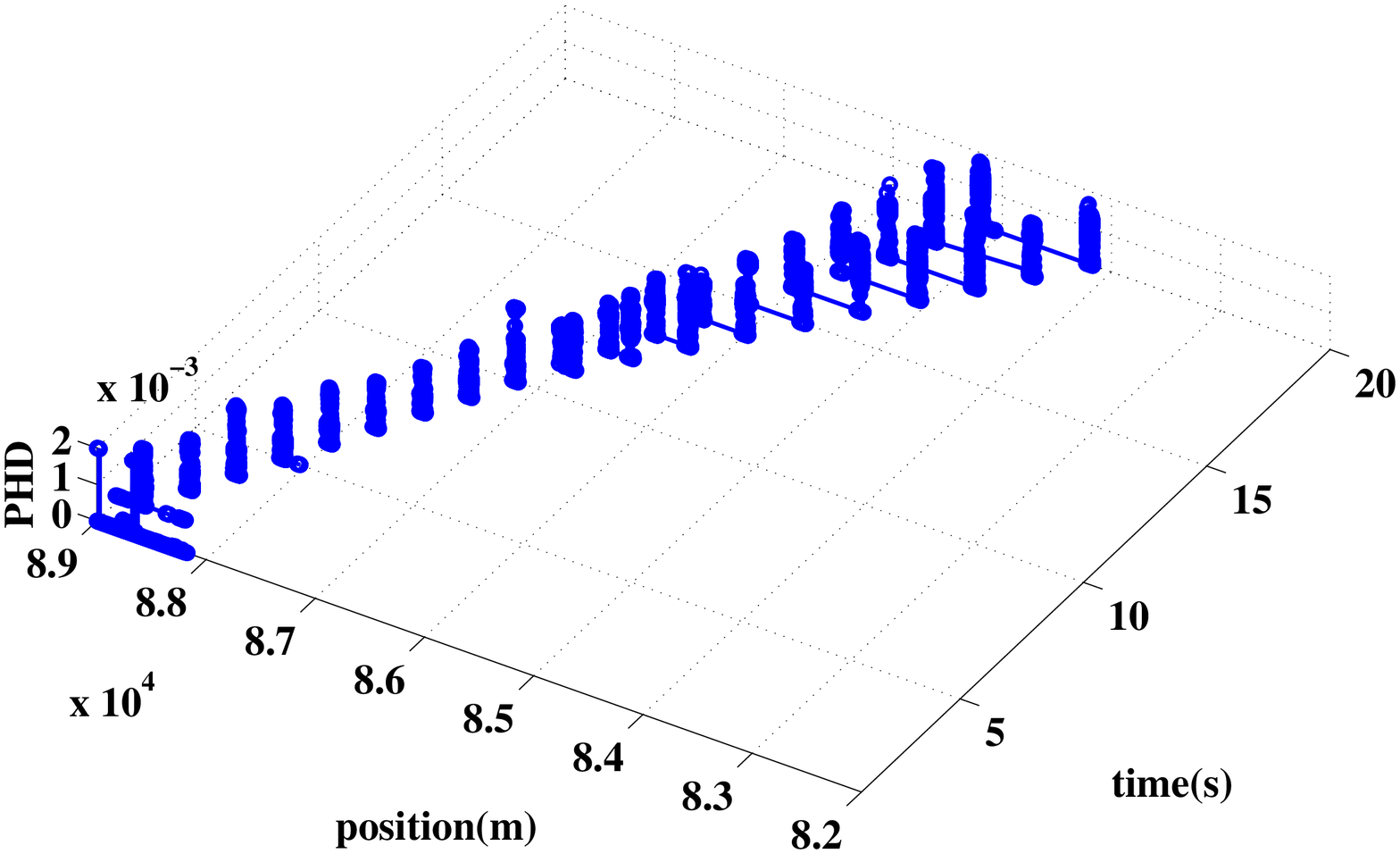}
\caption{PHD as a function of position and time step, using the
shrink-PHD filter.}
\label{fig:4}       
\end{figure}

\begin{figure}[!h]
  \includegraphics[width=0.75\textwidth]{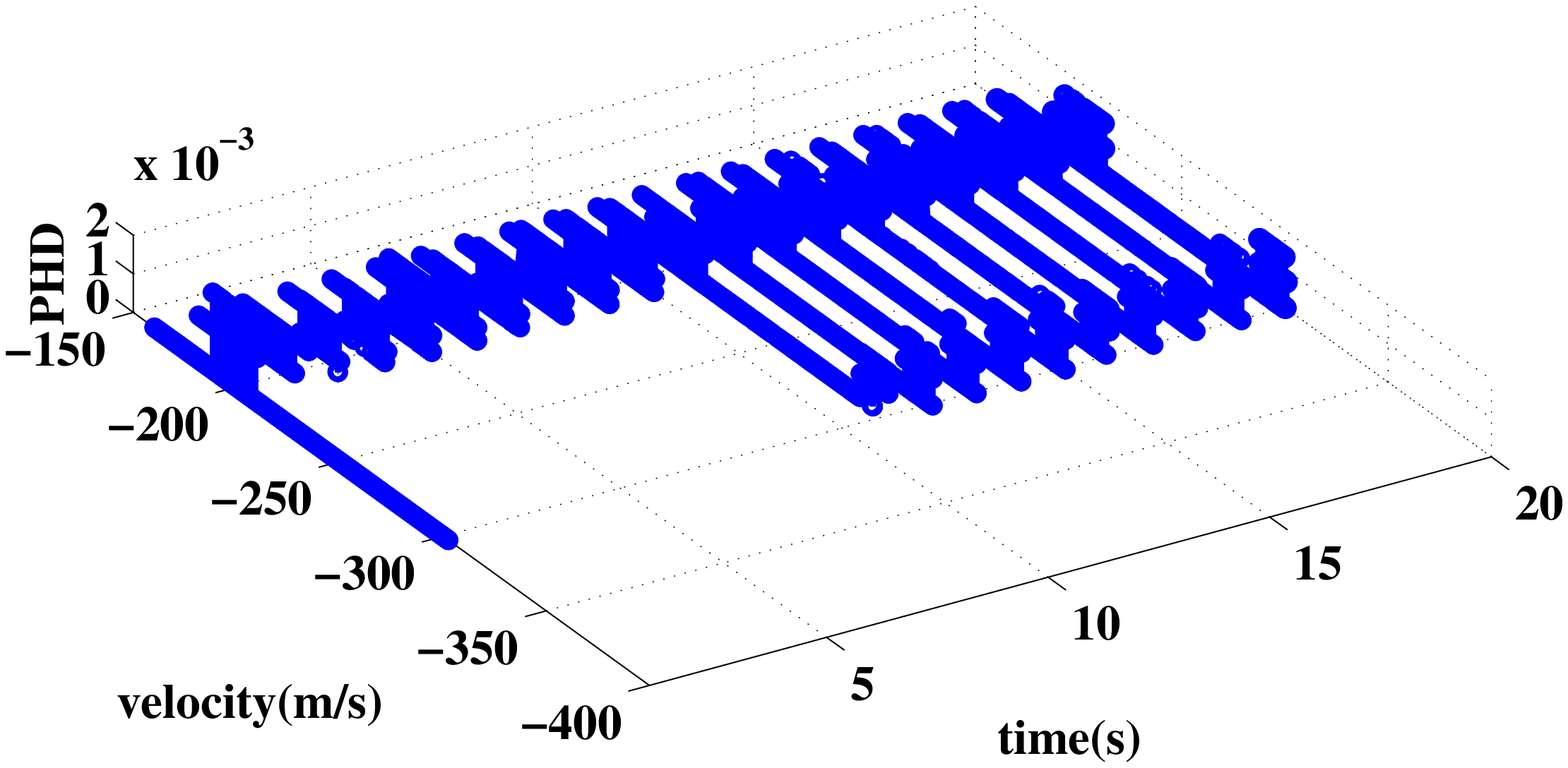}
\caption{PHD as a function of position and time step, using the
shrink-PHD filter.}
\label{fig:41}       
\end{figure}

In the first example, the SNRs of both targets are 9 dB. The
measurements are shown in Fig. 3. Tracking with the shrinkage-PHD
filter, we plot the PHD of the position and velocity versus time in
Fig. 4 and 5. It is shown that the PHD of targets can be integrated
over time by simultaneous detection and tracking. Observe that both
trajectories are automatically initiated and tracked. In step 3, the
first target can be tracked stably. At step 11, just 1 s after
target spawning, the second target is already initiated and tracked.

For the second example, we conducted simulations to compare the
performance of the shrinkage-PHD filter (shrinkage-PHDF) the PHD
filter (PHDF) \cite{Ref16}, and the classical multitarget particle
filtering method (MPF) \cite{Ref8} in tracking accurately and
consistently, and 50 Monte Carlo runs for the same scenario as used
in the first example were performed. The states of targets are
estimated by expectation-maximization (EM) algorithm. In addition,
the maximum possible number of targets is limited and known for MPF,
but unlimited and unknown for the PHD filter and the shrinkage-PHDF.

To measure the estimation quantitatively, the estimation errors in
terms of Monte Carlo averaged OSPA distance are shown in Figs. 6 and
7. For position estimation, the cut-off value c is five times the
size of the range cell. For velocity estimation, the cut-off value c
is two times the size of the velocity cell. It is shown that the
shrinkage-PHDF estimates target position and velocity on all tracks
with higher estimated precision than the MPF, because MPF requires a
modeling setup to accommodate varying numbers of targets, which is
very difficult to accomplish in reality.It also shows that the
performance of shrinkage-PHDF is better than PHDF, which is
discussed further below.

\begin{figure}[!h]
  \includegraphics[width=0.75\textwidth]{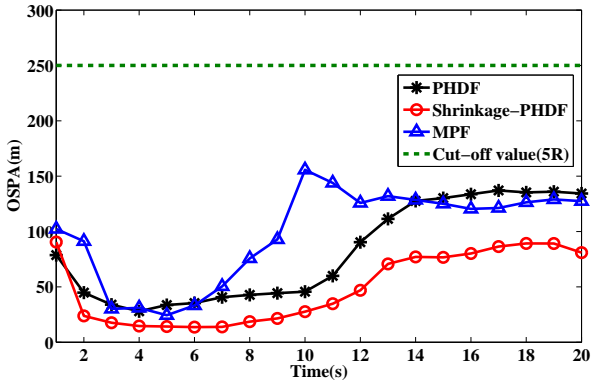}
\caption{Comparison of the estimation errors of position given by
the Monte Carlo-averaged OSPA miss distance with a measurement
resolution R=50 m.}
\label{fig:5}       
\end{figure}

\begin{figure}[!h]
  \includegraphics[width=0.75\textwidth]{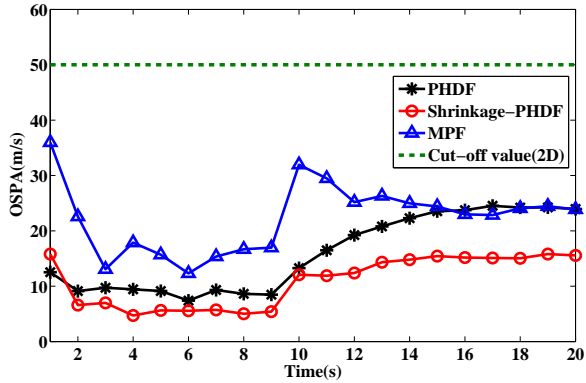}
\caption{Comparison of the estimation errors of the velocity given
by the Monte Carlo-averaged OSPA miss distance with a measurement
resolution D=25 m/s.}
\label{fig:6}       
\end{figure}

To compare the shrinkage-PHDF and the PHDF clearly, Figs. 8 and 9
show the comparison of the mean OSPA for varying SNRs. The SNR of
the targets is set to known values between 6 and 13 dB. When the
value is between 6 and 11 dB, the shrinkage-PHDF performs
significantly better than the PHDF. When the value is large than 11
dB, they perform similarly. This finding illustrates the limitation
of ordinary the PHDF for TBD: the PHD cannot sufficiently
approximate the multitarget posterior probability when the SNR is
low. The superiority of the shrinkage-PHDF takes advantage of the
shrinkage operation is indicated, especially for low SNR scenarios.
It is clear that for multitarget TBD, the shrinkage-PHDF performs
better than the PHDF and MPF, especially when the SNR is low.

\begin{figure}[!h]
  \includegraphics[width=0.75\textwidth]{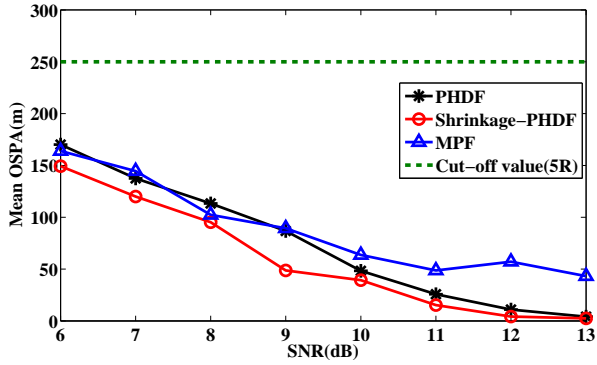}
\caption{Comparison of the estimation errors of position given by
the time-averaged OSPA miss distance with a measurement resolution
R=50 m for varying SNRs.}
\label{fig:7}       
\end{figure}

\begin{figure}[!h]
  \includegraphics[width=0.75\textwidth]{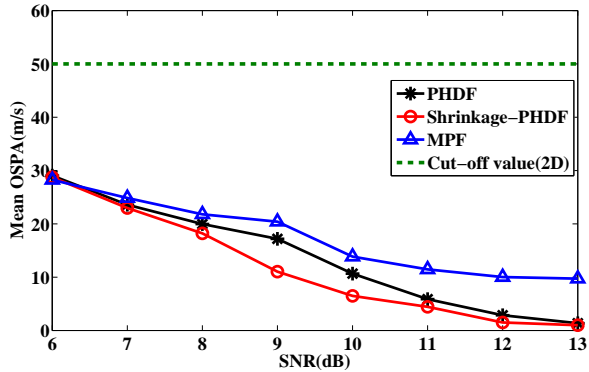}
\caption{Comparison of the estimation errors of the velocity given
by the time-averaged OSPA miss distance with a measurement
resolution D=25 m/s for varying SNRs}
\label{fig:8}       
\end{figure}

\subsection{Multiple targets with different SNRs}

In this example, the targets with different SNRs (the first target
is 8dB, the second target is 9dB and the last one is 10dB). At time
step1, the first target appears as in the first example. At step
7,the second target spawns form the first one, and the velocity in
the x-direction is -300m/s. The third target new-birth far from the
previous two targets at the time step 13 with the intital target
state ${\left[ {\begin{array}{*{20}{c}}
   {89000} & { - 250} & 0 & 0 \\
\end{array}} \right]^T}$. A comparison of the estimation errors of the position and
the velocity of the three algorithms is shown in Fig. 10 and 11. In
this scene, the targets vary not only the value of SNRs but also the
average number of targets. Because the SNR of the first target is
lower than which in section 6.2, the convergence velocity of the
estimation first target is slower. After time step 7, comparing
Figs. 6 and 10, Fig.7 and 11, the varying SNRs and the number of the
targets did not significantly influence the performance of the PHDF
and Shrinkage-PHDF. In this example, the performance of MPF is
better than that in the first example. The reason is the intensity
of the targets is the most important characteristic to differentiate
different targets for MPF method. In this way, the SNRs of all
targets should be known for MPF method.

\begin{figure}[!h]
  \includegraphics[width=0.75\textwidth]{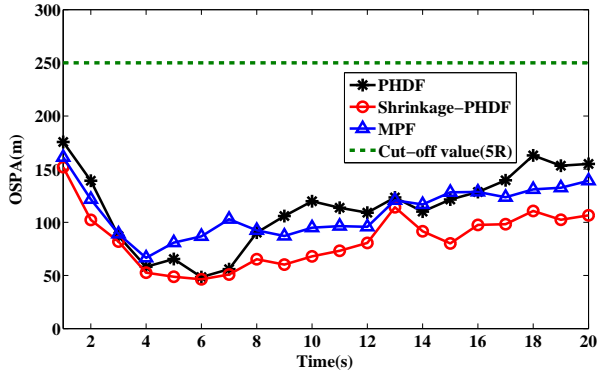}
\caption{Comparison of the estimation errors of the position given
by the Monte Carlo-averaged OSPA miss distance for multiple targets
with different SNRs (8dB, 9dB and 10dB).}
\label{fig:9}       
\end{figure}

\begin{figure}[!h]
  \includegraphics[width=0.75\textwidth]{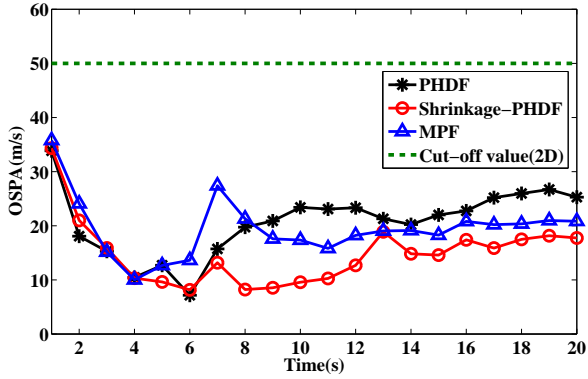}
\caption{Comparison of the estimation errors of the velocity given
by the Monte Carlo-averaged OSPA miss distance for multiple targets
with different SNRs (8dB, 9dB and 10dB).}
\label{fig:92}       
\end{figure}

\subsection{Multiple targets with unknown SNRs}

The PHD filter and the shrinkage-PHD filter are compared in Fig. 12
and 13. The sense is similar with that in the section 6.2, but the SNRs of the targets
are unknown and the SNR of the second target changes from 9dB to
10dB. It is obvious that the performance of both algorithms is
similar to what was presented in Figs. 6 and 7. After the 11th time step, the performance is even better than which in the section 6.2, because the value of the SNR of the second target is increased. Therefore, whether
the SNRs of multiple targets are known has not significant influence
on the performance of PHDF and Shrinkage-PHDF. In addition, MPF from
\cite{Ref8} does not work when the value of the SNRs of the targets
are unknown.

\begin{figure}[!h]
  \includegraphics[width=0.75\textwidth]{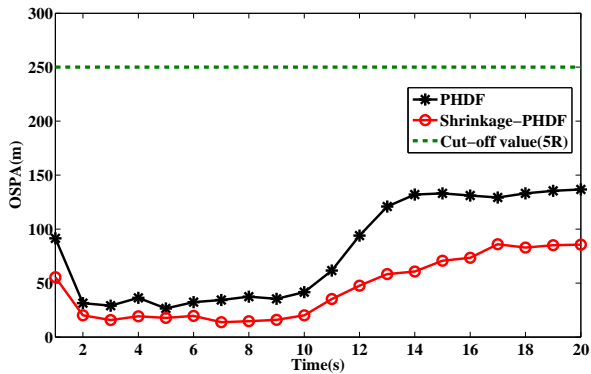}
\caption{Comparison of the estimation errors of the position given
by the Monte Carlo-averaged OSPA miss distance for multiple targets
with unknown SNRs (9dB and 10dB).}
\label{fig:101}       
\end{figure}

\begin{figure}[!h]
  \includegraphics[width=0.75\textwidth]{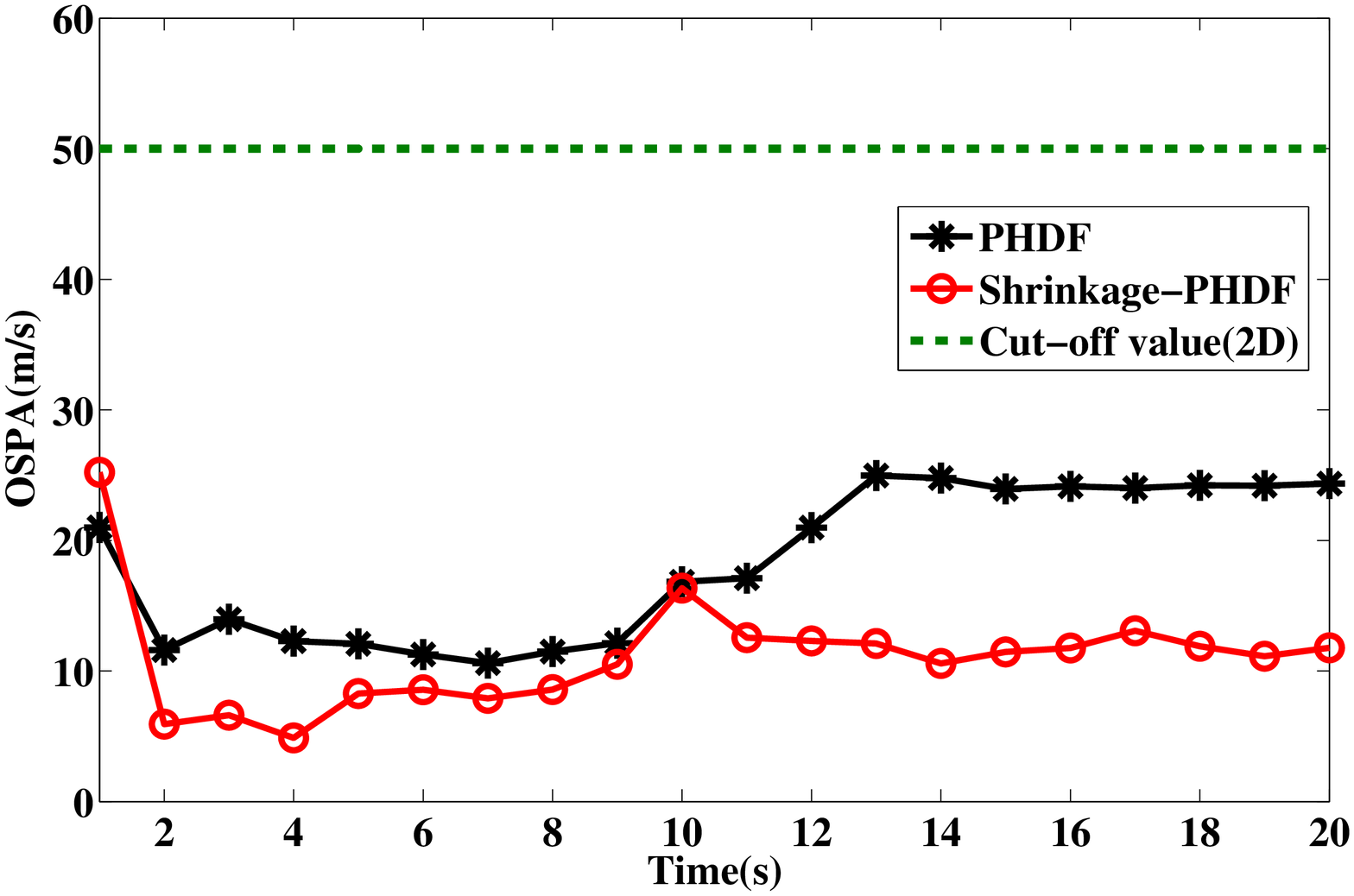}
\caption{Comparison of the estimation errors of the velocity given
by the Monte Carlo-averaged OSPA miss distance for multiple targets
with unknown SNRs (9dB and 10dB).}
\label{fig:10}       
\end{figure}

\subsection{A comparison of computation resources}

\begin{table}[!h]
\caption{Computation resources}
\begin{tabular}{llll}
\hline\noalign{\smallskip}
Algorithm & Shrinkage-PHDF & PHDF & MPF \\
\noalign{\smallskip}\hline\noalign{\smallskip}
Computation(Second)& 28.66 & 28.12 & 84.20 \\
\noalign{\smallskip}\hline
\end{tabular}
\end{table}
Table 3 shows the computation resources needed for the PHDF, the
shrinkage-PHDF and MPF. All simulations were run on a PC with an
Intel (R) Core TM2 Duo CPU E4500 @ 2.20 GHz processor. The time
listed is Monte Carlo-averaged, for a run in a scenario similar to
that given in Section 6.2. The amount of calculation time required
for the shrinkage-PHDF and the PHDF is much less than for MPF. The
calculated amount time required for the shrinkage-PHDF is slightly
greater than that required for the PHD filter.

\section{Conclusion}

In this paper, we used the RFS to model multitarget TBD measurements
and to design an efficient Shrinkage-PHD filter for multitarget TBD.
This filter accompanies with a shrinkage operation, and the optimal
parameter for the shrinkage operation is obtained by an optimization
procedure. Simulations show that the shrinkage-PHD filter takes very
little time to detect new targets, and it is sensitive to the
variation in the number of targets. Moreover, our algorithm can
track targets with high accuracy by taking advantage of the
shrinkage operation. In addition, there is no restriction on the
maximum number of possible targets or known value of SNRs of
targets. In a word, by the application of Shrinkage-PHD filter,
multiple targets could be tracked well in low SNR environments.

In the future research, a challenge to be explored for the
shrinkage-PHD filter will be determining how to model the
measurement of extended targets in the framework of FISST. If this
issue is resolved, it is believed that our new approach will be
useful in future radar systems using the TBD technique.




\end{document}